# Framework for the Hybrid Parallelisation of Simulation Codes

R.-P. Mundani[1], M. Ljucović[2] and E. Rank[1]

[1] Technische Universität München, Munich, Germany
[2] Western Michigan University, Kalamazoo, MI, USA

**Abstract**

Writing efficient hybrid parallel code is tedious, error-prone, and requires good knowledge of both parallel programming and multithreading such as MPI and OpenMP, resp. Therefore, we present a framework which is based on a job model that allows the user to incorporate his sequential code with manageable effort and code modifications in order to be executed in parallel on clusters or supercomputers built from modern multi-core CPUs. The primary application domain of this framework are simulation codes from engineering disciplines as those are in many cases still sequential and due to their memory and runtime demands prominent candidates for parallelisation.

## 1    Introduction

Typical simulation codes, especially from the engineering domains, entail huge memory and/or runtime requirements. Even the need for parallelisation is undoubted, in many cases parallelisation approaches lack in efficiency and scalability due to severe communication/synchronisation dependencies as well as data distribution and load balancing problems. Hence, a simple approach to derive parallel codes from existing sequential ones would be preferable.

On the other side, latest trends in hardware developments come up with increased intrinsic parallelism, i. e. a single CPU provides several cores to execute (blocks of) instructions in parallel. While such a multithreaded approach is quite elegant and (in most cases) easily to achieve, it usually does not scale with larger amounts of threads, lacks of sufficient support in complex task design (see [1] for instance) and according to [2] discards properties such as predictability and determinism. In case of hybrid parallelisation, i. e. the interplay of distributed and shared memory programming using MPI and OpenMP, resp., this becomes even worse, for instance due to insufficient thread safety within MPI calls (see [3]).

In this paper, we present a framework for hybrid parallelisation which is based on a strict job scheduling, where such a job can be anything from a complete program up to a single instruction. Those jobs – together with their dependencies on the results of other jobs – are defined by the user on any desired level of granularity. The difference and main advantage to 'classical' parallelisation is that the user does not need to care about communication and synchronisation of the single jobs as well as data distribution and load balancing which is all inherently carried out be the framework. This allows advancing from sequential to parallel codes with less effort as the complexity of the parallel program is (mostly) hidden from the user. In contrast to other approaches, such as [4], the framework is neither a new programming language nor introduces any new paradigm to be learned by the user.

The reminder of this paper is as follows. In chapter 2 we will introduce some basic definitions and the underlying idea of our framework while in chapter 3 we will describe its implementation and usage. In chapter 4 we will demonstrate the incorporation of a numerical example and also present the achieved results. Chapter 5 finally concludes this paper and gives a short outlook on future work.

## 2 Framework Concept

In order to describe the basic principles and concept of our framework, we first introduce our definition of a *job* which is essential for the further understanding and different from common task definitions that can be found in most literature, for instance [5].

### 2.1 Definitions

While an algorithm is usually defined as a set of instructions that have to be executed in a certain ordering, this doesn't say anything about its execution. It's the implementation that generally brakes down an algorithm to sequential code and destroys any parallel potential. Afterwards, it's often complicated to derive a parallel code from an exiting sequential one.

In order not to restrict ourselves in parallel code design, we start the other way round with parallel algorithms from which we derive the sequential ones, i. e. sequential algorithms can be treated as a special case of parallel algorithms. Hence, an algorithm consists of a set of *parallel segments S* that are executed in a given order where one segment $S_i$ might depend on the results of another segment $S_j$. Such a parallel segment contains a number of jobs $J$ that **all** can be executed at the same time – sufficient resources assumed – in arbitrary manner. A parallel segment is considered to be completed as soon as all jobs in that segment have terminated. A single job itself consists of a set of sequences of instructions $I$ where a single sequence $I_k$ is to be executed sequentially while **all** sequences – sufficient resources assumed – can be executed in parallel in arbitrary manner. A job is considered to be completed when the execution of all sequences in that job have terminated. Finally, the algorithm is considered to be completed when all parallel segments have terminated.

A hybrid parallel algorithm can now be defined in the following way. For a given set of parallel segments $S$ there exists at least one segment $S_i$ with a cardinality $|S_i| > 1$, i. e. it contains more than one job. Furthermore, there exists at least one job $J_k \in S_j$ with $|J_k| > 1$, i. e. it contains more than one sequence of instructions. If $S_i = S_j$ we would refer to as strict hybrid parallelism, otherwise ($S_i \neq S_j$) as loose hybrid parallelism. At this point we differ from the usual understanding of processes and threads. In the classical approach a job would correspond to an MPI process while a sequence of instructions would correspond to a thread—a constraint to the general idea. In our approach, a job can be both, MPI process and thread, which allows for a simpler scheduling of jobs and provides more flexibility concerning the underlying hardware. Nevertheless, several threads contained within one MPI process suffer from sufficient thread safety of MPI calls and, thus, have to be handled with care [3], for instance, gives a good overview over this set of problems.

## 2.2 Jobs

As already mentioned in the introduction, the idea of the framework is to hide as much complexity (i. e. communication, synchronisation, data distribution etc.) as possible from the user by providing him an interface that allows a simple specification of parallel jobs without the need to implement a whole parallel program. A job − in this sense − might consist of single instructions, blocks of instructions (loops, e. g.), functions, or an entire program. Further arguments to a job specification are the input and output data to be processed and returned, resp., by the job. The input data can be of any type but it has to be given in amount of *chunks*. This is necessary to allow the framework an automatic data distribution between all sequences (of instructions) within one job. By specifying one job's results as input to another job, any dependencies among those can be modelled, too.

Assume the following simple example where we want to find the maximum element *max* of a one-dimensional array *A*. Therefore, we define the input data as $k$ chunks $A_i$ of size $A/k$ and the output data as the maximum elements of each chunk. A job $J_1$ could now be specified as 3-tuple consisting of the function to be executed (let's say *search_max( )* for searching the maximum element within the given data), the amount $m$ of data chunks to be processed by this job, and finally the function's results. Similar a job $J_2$ can be specified for the remaining $k-m$ chunks. When running jobs $J_1$ and $J_2$ in parallel we are finally provided a vector of size $k$ with the maximum elements of all chunks. Hence, to obtain the global result, i. e. the maximum element of *A*, another job $J_3$ has to be specified, that (in this simple case) executes the same function *search_max( )* and takes as input the results of jobs $J_1$ and $J_2$.

We assume that the user application is a sequential code that now should be run in parallel. Therefore, the user needs to redefine his algorithm into parallel segments consisting of jobs. Generally speaking, this is a separation of the sequential code into parallel executable units which are grouped into parallel segments of jobs. The huge advantage of the framework is that it now takes all the cross process communication responsibilities from the user and that it further provides the user with the ability of harnessing distributed resources. In a first stage, the user defines how jobs are done while in a second stage he describes their mutual relationship and, thus, inherently the global scheduling of their execution. Once both parts are defined the user only needs to recompile the framework in order to get a hybrid parallel code to be run on a parallel machine. This code will – once executed – spawn processes (the term process is not restricted to MPI processes and also includes threads) and manages all tasks including data I/O and dynamic resource allocation.

## 3  Implementation

The framework distinguishes between two different types of running processes following the idea of a master-slave model. Nevertheless, our masters – so-called scheduler processes – contain the entire intelligence and job description and are started prior to the workers.

### 3.1  Job Scheduling

The scheduler processes are responsible for creating and managing the worker processes, assigning jobs to these workers, and managing all jobs' input and output parameters. Schedulers are of fixed size and stay 'active' during the entire execution of the program. Among all scheduler processes the one with $rank = 0$ in MPI_COMM_WORLD is the main or master scheduler, which is the only process that stores the complete algorithm description. All other schedulers ($rank > 0$) receive from this master information about which jobs they are responsible for and where they can access jobs' results that are under the supervision of other schedulers. While the master does not store any job related data except the job descriptions, all other schedulers store their jobs' results and further need to know how to assemble these results that might be requested as input arguments by any other job. Additionally, each scheduler ($rank > 0$) has a set of workers to whom it dispatches individual jobs for execution (see Figure 1).

On the other hand, worker processes are dynamically created (i. e. spawned during runtime) with the main objective to execute jobs assigned from their schedulers. Workers are *isolated* processes (w. r. t. other workers) that only know which job(s) to execute and where to receive/send the input/output data. They are intended to be memoryless, but they keep a copy of the input/output data of each job they execute until the responsible scheduler signals them the data is no longer required and can be deleted. In addition, workers can be completely detained from sending back any results, issuing a message instead that a job has successfully finished and the results are ready for being further processed. This allows reducing the communication overhead if any subsequent job takes these results as input arguments, for instance to be observed within iterative algorithms such as numerical solvers for linear equations systems, where one worker iteratively updates parts of the solution vector. The drawback of this approach is that in case a worker (due to some failure) has to be shut down, all results computed so far are lost and have to be re-computed.

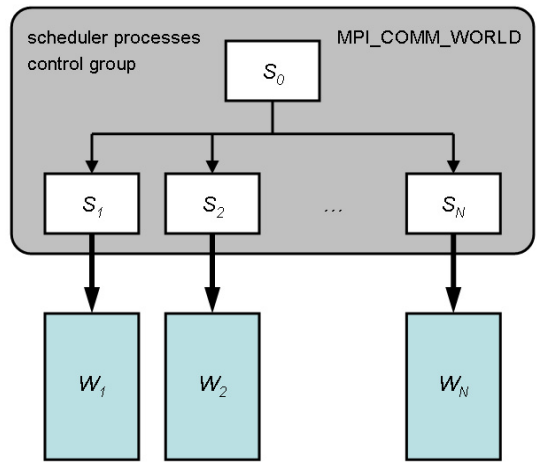

Figure 1: Scheduler control group with master $S_0$ that organises the program flow and its schedulers $S_1$ to $S_N$ that themselves assign and control job execution among their workers $W_i$

## 3.2 Registration of User Functions

There are two levels at which the user has to define his algorithm for a parallel execution. The first (logical) level describes the initial set of jobs (including dependencies) that the schedulers will assign to the worker processes while the second (physical) level describes the definition of user functions within workers. At the current stage of development we use the concept of 'fat' workers, i. e. there exists only one type of worker that includes all the functions implemented by the user. It is planed to extend the framework in a way to support 'slim' workers, i. e. that do a dynamic function loading during runtime and might also be specialised for different kind of hardware (GPUs, Cell Broadband Engine etc.). A typical worker process is given by the following code sample that shows how users can register their functions to workers before recompiling the framework.

```
AbstractWorker *worker = new GeneralWorker()
worker->init()

worker->add_function(user_function_1)
worker->add_function(user_function_2)
        ...
worker->add_function(user_function_n)

worker->run()
worker->finalize()
```

As each worker contains a number of user functions, it is within the user's responsibility to register these functions which should have the following signature.

```
void function_name(FunctionData *input, FunctionData *output)
```

Each *FunctionData* consists of a number of data chunks where such a chunk represents one consecutive memory location storing some quantity of an MPI data type also including user defined ones. In the latter case, the user needs to further supply a definition function which will be called during the initialisation phase on both workers and schedulers. Data chunks are arrays containing *n_elem* elements of a single MPI data type and are defined in the following way.

```
DataChunk(MPI_type datatype, int n_elem, void *data)
```

Here, the last argument is a pointer to the memory location where the real data is stored. For performance reasons, *DataChunk( )* copies the pointer to the data instead the data itself. Hence, the user must not delete the data whose pointer was given to *DataChunk( )* which – in turn – is responsible for deleting the data in case it is no longer needed. Finally, the following example illustrates how the user can access the input data and store the computed results (to the location indicated by *output*) within a user defined function.

```
void square(FunctionData *input, FunctionData *output)
{
   int a = * (int *) (input->get_data_chunk(0)->get_data())
   int *result = new int
   *result = a*a
   output->push_back(new DataChunk(MPI_INT, 1, result)
}
```

Figure 2 puts everything together and depicts the flow control of the framework in case of two schedulers. The master selects one of the available jobs for execution and tells either $S_1$ or $S_2$ to take care about the job and its results for a later processing. Again, the master does not store any results which are solely retained by the schedulers

### 3.3 Job Definitions

Input to the master scheduler is (so far) a plain text file containing a sequence of parallel segments separated by semicolons where each sequence consists of a comma-separated list of jobs. Since each job must know what to do, its definition takes four integer values as arguments:

- function identifier (a number as defined within worker process),

- number of threads needed (0 indicates as many threads as available cores of the underlying CPU; any number > 0 indicates the exact amount of threads),

- number of data chunks to be processed (0 if none or $J_i$ [$C_1$, $C_2$,…,$C_N$] indicating to take data chunks $C_k$ from the results computed by job $J_k$),

- *true/false* (optional clause (default *false*) – job will not send back results to its scheduler).

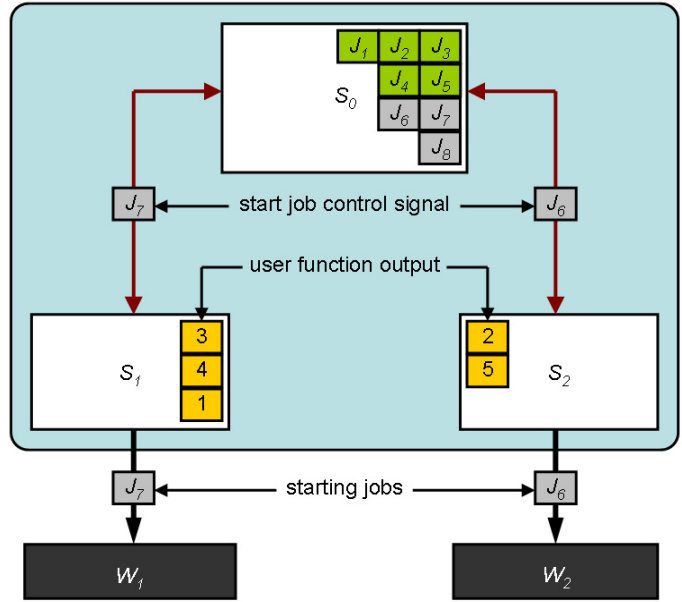

Figure 2: Flow control of the framework – the master $S_0$ selects jobs not processed so far and assigns them to the schedulers (here $S_1$ and $S_2$) for execution on the corresponding workers $W_i$

A sample input file might then look as follows.

```
J1(1,0,0), J2(2,1,0);
J3(2,2,R1[0..5],true), J4(2,2,R1[5..10],true), J5(3,0,R1 R2),
    J6(4,0,R1 R2);
J7(5,1, R2 R3 R4 R5);
```

Here, first two jobs $J_1$ and $J_2$ are defined to be executed in parallel (recall that all jobs of one parallel segment can be executed at the same time). Job $J_1$ calls user function 1 with the maximum amount of threads and takes no input arguments while job $J_2$ calls user function 2 with a single thread and also no input arguments. Hence, the framework starts two workers $W_1$ and $W_2$ and assigns each of them one of the two jobs $J_1$ and $J_2$. Once $J_1$ and $J_2$ have finished they return their results to the framework which continues with the execution of $J_3$ to $J_6$. While $J_3$ and $J_4$ only take some chunks of $J_1$'s results ($R_i$ indicating results if job $J_i$), $J_5$ and $J_6$ take both the entire results of $J_1$ and $J_2$. At this point the framework also has to spawn more workers in order to assign all four jobs to available workers in way thus they can be executed in parallel. As jobs $J_3$ and $J_4$ both intend to call user function 2 with two threads each, the framework could exploit this by assigning both jobs to the same worker (if running on at least a 4-core CPU). This is another advantage of the framework as it can optimally utilise the available resources. Finally, once job $J_7$ finishes we consider the algorithm to be completed and the framework shuts down.

Another important issue is, that during runtime each job can add a finite number of new jobs to the current or following parallel segments. This is necessary in case of iterated executions as shown for a Jacobi solver in the next section.

## 4    Example and Results

In order to test the framework we have implemented a parallel Jacobi solver for linear equation systems $A \cdot x = b$, which is one of the core parts of any numerical simulation code. The typical sequential code looks as follows.

>   **while** *res* > ε **do**
>       **for** $i \leftarrow 1$ **to** $N$ **do**
>           compute update $y_i \leftarrow b_i - \sum_{j \neq i} \alpha_{ij} \cdot x_j$
>       **od**
>       apply all updates $x_i \leftarrow (x_i + y_i)/\alpha_{ii}$
>       compute residual *res*
>   **od**

Nevertheless, this simple algorithm bears some severe problems when it should be executed in parallel via the framework. Its main part consists of computing one iteration of the update vector $y$ and will be defined as job $J_1$. Further jobs comprise applying the updates and computing the residual ($J_2$) as well as the outer loop checking for convergence ($J_3$). While $J_1$ and $J_2$ are straightforward and their sequences can easily be run in parallel, job $J_3$ causes a problem as the iterated execution of $J_1$ and $J_2$ is difficult to express as job definition according to the description given above. Hence, there was the need to allow jobs themselves the creation of new jobs. For the example of the Jacobi algorithm, job $J_3$ evaluates the input retrieved from $J_2$ and – if necessary – enforces the newly execution of $J_1$ and $J_2$ by adding them back again to the master scheduler.

What, at a first glance, sounds inefficient and cumbersome has been tested for different problem sizes and amount of processes compared to a 'tailored' solution, i. e. an efficient (solely) MPI implementation of the Jacobi solver. Even it was obvious that the framework cannot be competitive with such a tailored solution, the achieved results are very promising and vary (mean value) around 10 % from the runtime of an efficient MPI implementation (see Figure 3). Nevertheless, the runtimes from the framework are based on a sequential code that neither has been modified very much nor required lots of effort for its parallelisation—and those were the main objectives of our framework.

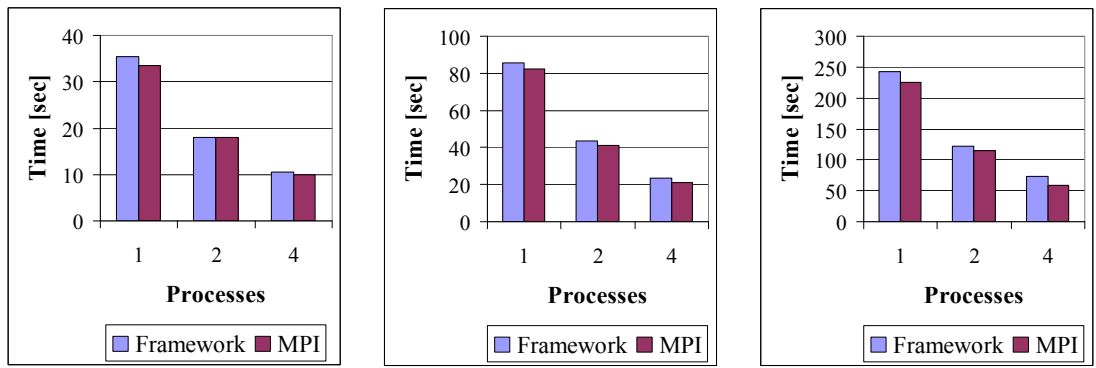

Figure 3: Achieved results (500 iterations) of a parallel Jacobi solver for different sizes (2709 × 2709 left, 4209 × 4209 middle, 7209 × 7209 right) with the framework and a corresponding MPI implementation

# 5   Conclusions

We presented a framework for hybrid parallelisation that allows the user to simply incorporate any sequential algorithm in order to be executed in parallel. The major benefits of this framework are due to its encapsulation of all underlying MPI and OpenMP complexity from the user as well as its automated management of available resources for an optimal execution of the user defined jobs. Next steps will comprise the testing of the framework with more complex simulation codes (FEM-based structure simulation, e. g.), the implementation of basic monitoring and fault tolerance properties, and its application on different hardware such as GPUs or the Cell Broadband Engine.